\documentclass[lettersize,journal]{IEEEtran}

\ifCLASSOPTIONcompsoc
  \usepackage[nocompress]{cite}
\else
  \usepackage{cite}
\fi

\ifCLASSINFOpdf
\else
\fi

\usepackage[utf8]{inputenc} 
\usepackage[T1]{fontenc}    
\usepackage{hyperref}       
\usepackage{url}            
\usepackage{booktabs}       
\usepackage{amsfonts}       
\usepackage{nicefrac}       
\usepackage{microtype}      
\usepackage{amsmath}
\usepackage{graphicx}
\usepackage{multirow}
\usepackage{multicol}
\usepackage{subfigure}

\usepackage{algorithm}
\usepackage[noend]{algpseudocode}
\usepackage{xcolor}
\usepackage{arydshln}
\usepackage{enumitem}
\usepackage{thm-restate}
\usepackage{amsmath,amsfonts,bm,amssymb,amsthm}

\begin{document}

\title{Fairness Attacks on Recommender Systems}

\author{Yanan Wang,
        Yong Ge
\IEEEcompsocitemizethanks{
\IEEEcompsocthanksitem 
Yanan Wang is with Department of Information Systems and Operations Management, College of Business, The University of Texas at Arlington. Yong Ge is with Department of Management Information Systems, Eller College of Management, the University of Arizona.
\protect\\E-mail: yanan.wang@uta.edu, yongge@arizona.edu
}
}

\IEEEtitleabstractindextext{
\begin{abstract}
The unfairness of recommender systems has become a topic of concern due to its significant social and ethical implications. Although existing works have shown the effectiveness of attacks on the performance of recommender systems (e.g., promotion and demotion attack), the study of fairness attacks on recommender systems remains largely under-explored. To this end, we propose a novel structure-aware reinforcement learning-based fairness attack method designed to exacerbate the unfairness of target recommender systems. Specifically, we first employ a graph-based structure encoder to model the structural dependencies among the generated fake user-item interactions and the original user-item interactions. Then, we model the sequential dependency of the injected fake items using a recurrent neural network. Based on the learned structure-aware and sequence-aware representations of the fake user and item, the item selection policy attentively decides the next injected fake item. Since the target recommender system may employ fairness-aware training and leverage the user's sensitive attribute information, such as gender, we further designed a gender selection policy to decide the gender of the entire fake user profile. Both the item selection and gender selection policy are learned jointly in our proposed method. Finally, experimental results on four types of target recommendation models and two real-world datasets demonstrate the effectiveness of the proposed attack method in exacerbating the unfairness of recommender systems.
\end{abstract}

}

\maketitle

\IEEEdisplaynontitleabstractindextext
\IEEEpeerreviewmaketitle

\section{Introduction}\label{Sec:intro}
Recommender systems aim to provide personalized recommendations to cater to user interests, thereby mitigating information overload. They are typically trained using collected historical user-item interaction data and, post-training, predict items potentially interesting to users. Due to its significance, recommender systems are employed across various domains, including e-commerce \cite{covington2016deep}, education \cite{urdaneta2021recommendation}, and employment \cite{kenthapadi2017personalized}.

Although recommendation services considerably enhance daily life, contributing to significant economic and social impacts, they face an escalating challenge: unfair recommendations, wherein recommenders systematically and unfairly discriminate against specific user groups. Such unfairness has been observed in online learning recommendation services, where STEM lectures are disproportionately recommended to minority groups \cite{yao2017beyond}, in online job recommendations, favoring male candidates over female ones for technical roles \cite{kochling2020discriminated}, and in online rental recommendations, where certain apartment types are not suggested to minority groups like single parents \cite{spinks2019contemporary}. The pervasiveness of this unfairness results in inequitable information services for end-users, leading to social injustice in education and career development, and legal and ethical issues for system providers \cite{tobin2019hud}. To tackle this, researchers and practitioners have developed fairness-aware recommenders focused on addressing the inherent bias in consumer-generated ratings to mitigate recommendation unfairness \cite{beutel2019fairness, bose2019compositional,wang2023survey,ekstrand2018all}.

Most recommender systems, e.g., YouTube, are accessible to external users who can create accounts, input ratings, and access most historical ratings provided by other users. This openness allows adversaries to inject intentionally generated fake ratings, influencing the underlying algorithm post-training, and leading to recommendations desired by the adversaries. Previous studies have documented evidences of such adversarial attacks promoting or demoting products in recommendation lists across different recommender systems \cite{tang2020revisiting,yang2017fake}. Hence, it is plausible for adversaries to launch similar attacks to harm recommendation fairness, exacerbating the recommender system's unfairness issue. Various parties are also motivated to perform such adversarial fairness attacks on recommender systems. For example, opponents of a recommendation platform may attempt to damage its reputation and trustworthiness to gain a competitive market advantage or even embroil the platform owner in legal trouble. Despite the high likelihood and strong motivation, to our knowledge, such adversarial fairness attacks on recommender systems remain underexplored.

\begin{figure}[t]
    \centering
    \begin{minipage}[t]{0.45\textwidth}
        \centering
        \includegraphics[width=\textwidth]{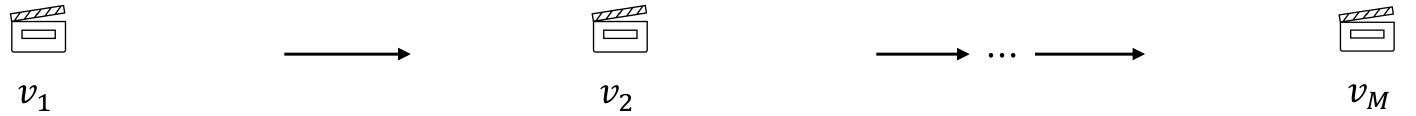}
        \\[1ex]
        \small (a) Fake item selection in existing RL-based recommendation performance attack methods
    \end{minipage}
    \hfill
    \begin{minipage}[t]{0.45\textwidth}
        \centering
        \includegraphics[width=\textwidth]{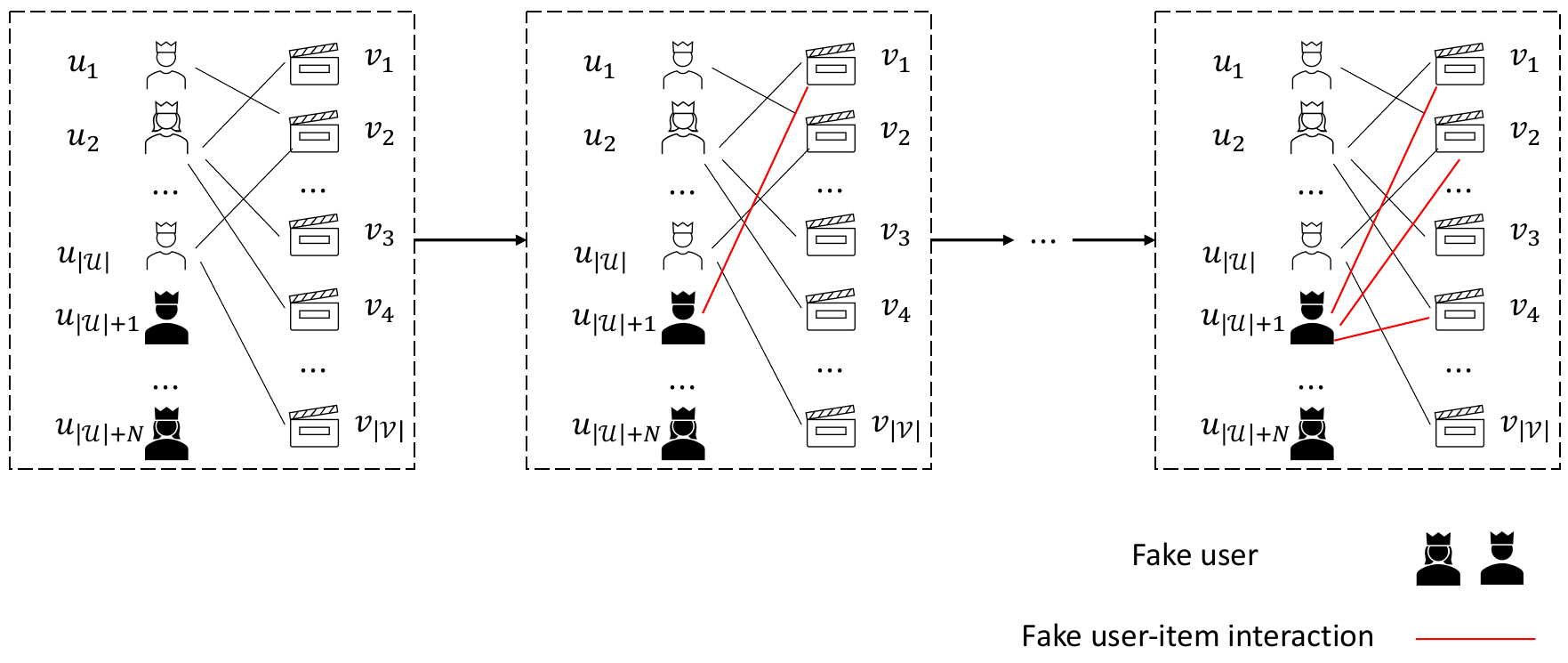}
        \\[1ex]
        \small (b) Fake item selection in the proposed structure-aware fairness attack method
    \end{minipage}
    \caption{Comparison of fake item selection strategies.}
    \label{fig:intro}
\end{figure}

From a cybersecurity perspective, this motivates a \emph{red-teaming} mindset for recommender fairness. In the cat-and-mouse dynamics between defenders and adaptive adversaries, robust evaluation requires \emph{aggressive} attack simulation. If attacks used in evaluation are weak, we may substantially underestimate how much a recommender's fairness can deteriorate, leading to a false sense of robustness \cite{khaleel2024network}. Therefore, developing advanced fairness attack algorithms is a prerequisite for proactively identifying vulnerabilities, evaluating mitigation mechanisms, and ultimately improving the resilience and accountability of the fairness of recommendation services.

To address this crucial research gap, we examine the problem of fairness attacks on recommender systems intending to harm their fairness. Designing a fairness attack algorithm presents several challenges. Firstly, due to the existing fairness issue in recommendations, the target recommender system may employ fairness-aware training to improve recommendation fairness. Such potential fairness-aware recommendations also leverage sensitive attribute information in the training data, which, in this paper, is the gender attribute. Therefore, besides deciding on selecting fake items in the recommendation performance attack setting \cite{tang2020revisiting}, fairness attackers also need to determine the gender attribute information of each fake user profile to maximize the fairness attack performance. Jointly deciding the fake items and gender attributes of each fake user profile is the first challenge. Secondly, during the fake user profile generation process, attackers must decide each item in the fake user profile and each gender attribute in the created fake user profile. This multi-step decision process naturally aligns with reinforcement learning \cite{sutton2018reinforcement,silver2016mastering}, which is well-suited to solve such decision processes. A few works have applied reinforcement learning (RL) to attack the recommendation model's performance \cite{song2020poisonrec, zhang2020practical}. However, as shown in Fig. \ref{fig:intro} (a), these existing RL-based recommendation performance attack methods overlook the structural dependency among the injected fake user-item interactions and the original user-item interactions. Therefore, modeling the structural dependency is the second challenge.

To overcome these challenges, we propose a novel Structure-aware Reinforcement Learning based Fairness Attack (SRLFA) method to jointly learn the fake item selection and gender attribute selection strategy. Specifically, the user-item interactions can be naturally viewed as a bipartite graph. As depicted in Fig. \ref{fig:intro} (b), the fake user nodes are added to the original user-item interaction graph. Each subsequent injected fake user-item can be viewed as adding a new fake edge to the original user-item interaction graph. Therefore, both the added fake user nodes and fake user-item interaction edges alter the structure of the original user-item interaction graph. To model the structural dependency among the fake user-item interaction and the original user-item interaction graph, we utilize a graph-based structure encoder to learn the structure-aware representation of each fake user and fake item. As each fake item is decided sequentially, we further model the sequential dependency among the generated fake items using a recurrent neural network. Based on the learned structure-aware and sequence-aware representations, the item selection policy attentively decides the selected fake item at the current time step. To decide each fake user profile's gender attribute, we further learn a gender selection policy based on the structure-aware and sequence-aware representation on the user profile level. In practice, the target recommender system is often unknown to the attacker. To optimize the fake user profile based on feedback, we employ a local surrogate recommender. This surrogate is designed with fairness-aware training, emulating potential fairness considerations in the target recommender systems. Both the item selection policy and gender selection policy are learned jointly using reinforcement learning.

We conduct comprehensive experiments to evaluate our developed method. We adopt two representative target recommendation models and assess the fairness attack performance on two real-world datasets. To simulate the potential fairness-aware training of the target recommendation model, we further evaluate the fairness attack performance on the target recommendation models incorporating fairness-aware training. Experimental results demonstrate the proposed fairness attack method's superiority compared to baseline attack methods. In summary, this paper's main contributions are as follows:
\begin{itemize}
\item We explore the scarcely investigated problem of fairness attacks on recommender systems, which can have a profound social impact.
\item We propose a novel structure-aware reinforcement learning based fairness attack method, modeling the structural dependency among the generated fake user-item interaction and original user-item interaction graph.
\item In the proposed method, the fake item and gender attribute of the fake user profile are jointly learned in an end-to-end manner using reinforcement learning.
\item Extensive experiments demonstrate the effectiveness of the proposed method in exacerbating the unfairness of original recommender systems and fairness-aware recommender systems, which servers as an important foundation for studying the vulnerability of recommendation fairness.
\end{itemize}

The remainder of this paper is organized as follows: Section 2 offers a brief review of the existing literature on attacks against recommender systems, machine learning fairness, and fairness-aware recommender systems. Section 3 introduces the
concept of a fairness attack and the associated threat model, including the attacker’s goal, knowledge, and capability. In Section 4, we elaborate on our proposed method for conducting fairness attacks. Section 5 discusses the results of our experiments. Finally, we concludes the article in Section 6.

\section{Related Work}
In this section, we briefly review the related works from two categories: attacks on recommender systems, attacks on machine learning fairness and fairness in recommender systems.
\subsection{Attacks on Recommender Systems}
Due to openness of the recommender systems, previous researches have shown that it is possible to attack the recommender systems \cite{yang2017fake, tang2020revisiting, wu2021triple,lee2012shilling}. The attackers can inject the fake user profiles into the training data of the target recommendation systems and the recommendation results can be manipulated after the recommender system get retrained with these malicious data. There are two types of attacking goals: promotion attack and demotion attack \cite{yang2017fake}. Promotion attack aims to promote a set of target items while demotion attack aims to demote a set of target items.

Existing attacking methods can be classified into two categories: white-box attack and black-box attack. White-box attack \cite{li2016data,christakopoulou2019adversarial,fang2020influence,fang2018poisoning,yanan2024attack} assumes the parameters and model type of the target recommender system is known to the attackers. This kind of assumption is usually unpractical in the real world as it is hard for the attackers to have a full access to the target recommender system. While the black-box attack does not assume the knowledge of the target recommender system and regard the target recommender system as a black box. To generate the fake user profiles, existing black-box attack methods either adopt a local surrogate recommender or query the target recommender \cite{tang2020revisiting, song2020poisonrec,zhang2021data}. Among the attack strategies used by the black-box attack methods, early shilling-based attack methods \cite{chirita2005preventing} adopt heuristic rules to manually create the fake user profiles and the attack performance is limited. Machine learning-based attack methods are later proposed to improve the attack performance. Revisit method \cite{tang2020revisiting} updates the fake user profile using the gradient of the adversarial loss, where the adversarial loss is related to the likelihood of target items in the recommendation list. Generative adversarial neural network has also been used to generate the fake user profiles to approximate the ratings of real users \cite{lin2020attacking}. Recently, a few works formulate the fake user profile generation as a multiple step decision making problem and adopt reinforcement learning to solve this problem \cite{song2020poisonrec, zhang2020practical}. The limitations of existing RL-based recommendation performance attack methods are discussed in Section \ref{Sec:intro} as well as Fig. \ref{fig:intro}.

However, all these works focus on attacking the performance of recommender systems, i.e., promotion or demotion attack. While in this paper, we focus on the fairness attacks on recommender system, which aims to decrease the fairness of the recommender system and has a profound social impact. Moreover, we design a hierarchical fairness attack policy to decide the selected items and gender of each fake user profile. The developed fairness attack policy further incorporate user-item interaction graph to model the structural influence of each injected fake user-item interaction, which is ignored in previous works.
\subsection{Attacks on Machine Learning Fairness}
Due to the importance of the fairness of machine learning algorithms, a few recent works study the vulnerability of machine learning fairness by design attack methods. Gradient-based and influence-based attack has been developed to decrease the fairness of machine learning classification algorithms \cite{mehrabi2021exacerbating,solans2020poisoning}. Gradient-based attack method \cite{solans2020poisoning} derive the gradient of the adversarial loss that maximize classification disparities. Influence-based attack 
\cite{mehrabi2021exacerbating}method estimates the influence of the poisoned sample on the unfairness of the classification algorithm using the influence function \cite{koh2017understanding} and then update the poisoned sample using the estimated influence to maximize the unfairness. 

Different from these attacks on the fairness of machine learning classification algorithms, this paper aims to study the fairness attacks on recommender systems, which is a fundamental different task. Moreover, in contrast to the gradient and influence-based fairness attack method \cite{mehrabi2021exacerbating,solans2020poisoning}, our fairness attack method is based on reinforcement learning by utilizing the structure information of user-item interactions.

\subsection{Fairness-aware Recommender Systems}
Recent advancements highlight the growing importance of fairness in recommender systems due to its significant societal implications \cite{wang2023survey}. A variety of fairness-aware recommender systems have been developed to enhance recommendation fairness \cite{wang2023survey,rastegarpanah2019fighting,li2021user}. These fairness-centric recommendation techniques can be categorized into three main categories: pre-processing, in-processing, and post-processing \cite{li2023fairness}. Pre-processing methods \cite{rastegarpanah2019fighting,ekstrand2018all} focus on modifying the training data to improve the fairness of recommendation model. In-processing methods \cite{yao2017beyond,zhu2018fairness} change the training process of recommendation models to reduce unfairness. A notable example is the regularization-based approach where a fairness regularization term is incorporated into the matrix factorization loss function to enhance fairness \cite{yao2017beyond}. Lastly, post-processing methods modify the model output to improve the fairness \cite{li2021user,patro2020fairrec}. 

In this study, we not only explore fairness attacks on conventional target recommendation models but also examine such attacks on models that integrate fairness-aware training. Specifically, we choose the fairness regularization method as the representative fairness-aware training techniques.
\section{Problem Statement}\label{sec:problem_state}
In this section, we first briefly introduce the recommendation task and then provide the definition of recommendation fairness. After that, we formally define the problem of fairness attack by illustrating the attacker's goal, attacker's knowledge and attacker's capability.

A recommender is essentially an algorithm trained on historical user-item interaction data. This interaction data can be represented by the matrix \(X \in \mathbb{R}^{|U| \times |V|}\), where \( \mathcal{U}=\{u_1, \ldots, u_{|\mathcal{U}|}\} \) and \( \mathcal{V}=\{v_1, \ldots, v_{|\mathcal{V}|}\} \) denote the sets of users and items in the recommender system, respectively. An interaction (e.g., click, purchase) between the \(i\)-th user and the \(j\)-th item results in a non-zero matrix entry. This paper emphasizes user's implicit feedback, which is often more accessible in real-world recommendation scenarios. Based on the historical user-item interaction data $X$, the goal of the recommender system is to provide recommendations to match user's interest.

Apart from evaluating the overall effectiveness of recommendations, the fairness in recommendations has received significant attention recently \cite{wang2023survey,rastegarpanah2019fighting,li2021user}. In the group fairness definition, users can be categorized into different groups according to the sensitive attribute such as gender and race. To adhere to this fairness principle, recommendation systems should yield comparable performance across distinct user groups. For our analysis, we consider two groups of users denoted as $G_1$ and $G_2$ (e.g., male and female users). We prioritize gender as the primary sensitive attribute in this study. The performance disparity (unfairness) among different gender groups is defined as follows:

\begin{equation}\label{eq:unfair_metric_define}
    Unfair_{\mathcal{F}}=\left|\frac{1}{\left|G_1\right|} \sum_{u \in G_1} \mathcal{F}\left(u\right)-\frac{1}{\left|G_2\right|} \sum_{u \in G_2} \mathcal{F}\left(u\right)\right|,
\end{equation}
where $\mathcal{F}$ is the recommendation performance metric such as NDCG, Precision and F1 score.

Based on the definition of the $Unfair_{\mathcal{F}}$, given a target recommendation model, the threat model is defined as follows:

\textbf{Attacker's Goal}: The goal of the attacker is to compromise the fairness of the target recommendation model by increasing the unfairness metric: $Unfair_{\mathcal{F}}$.

\textbf{Attacker's Knowledge}:
In our assumption, attacker does not have the knowledge of the parameters and models used in the target recommender system, which is the black-box attack setting. Due to the openness of the recommender system (e.g., Amazon,YouTube), it's reasonable to assume the attackers have the access to full or partial training data used by the target recommender system \cite{tang2020revisiting,zhang2020practical}.

\textbf{Attacker’s Capability}:
The attacker has the capability to manipulate numerous fake user profiles and inject them into the target recommendation system. Given practical constraints on the attacker's resources, we posit that the count of these fake user profiles is $N$, with each profile encompassing $M$ interactions. each fake user profile integrates the sensitive attribute, represented by $g_{u_i}$, along with a list of interacted items, expressed as $\hat{x}_{u_i} = \{g_{u_i}; v_0, \cdots, v_{M - 1}\}$.

\section{Method}
During the fake user profile generation process, the attacker decides each fake item sequentially and update the attacking strategy by receiving feedback from the recommendation model, which naturally fits to the reinforcement learning setting \cite{sutton2018reinforcement,silver2016mastering}. Therefore, we formulate this sequential decision making process as a Markov Decision Making (MDP) process. In this section, we first provide the formulation of the MDP process. Then, we provide an overview of the proposed method. After that, we detail each component of the framework. Finally, we provide the training algorithm of the proposed model.
\subsection{MDP Formulation}
We formulate the fake user profile generation as an MDP process. The MDP is defined as a tuple $(\mathcal{S}, \mathcal{A}, \mathcal{P}, \mathcal{R}, \gamma)$, where $\mathcal{P}$ is the state transition probability, $\gamma$ is the discount factor and the rest is listed as follows:
\begin{itemize}
\item \textbf{Action $\mathcal{A}$}: There are two types of actions when generating each fake user profile: the item selection action and gender selection action. The selected item at time $t$ is $a^{item}_t \in \mathcal{V}$. The gender selection action is $a^g_{u_i} \in \{0, 1\}$, where 0 and 1 represent the male and female genders, respectively. The gender action is decided every $M$ steps, where $M$ is the number of selected items in each fake user profile.
\item \textbf{State Space $\mathcal{S}$}: $s_{t} \in \mathcal{S}$ is the previous generated fake users and selected items before time $t$. $s_{t} = \{u_{|\mathcal{U}| + 1}, a_0^{item}, a_1^{item}, \cdots, a_{M - 1}^{item}, u_{|\mathcal{U}| + 2}, \cdots, a_{t - 1}^{item}\}$.
\item \textbf{Reward $\mathcal{R}$}: As the goal of the attacker is to increase the unfairness of the target recommender system, the reward is defined as the $Unfair_{\mathcal{F}}$ in Eq. \ref{eq:unfair_metric_define}. Specifically, we adopt the unfairness metric $Unfair_{F1}$. Due to the black-box attack setting, the attack has no knowledge about the target recommendation model. Therefore, we adopt the surrogate recommender to provide reward feedback. The reward is obtained every $\tau$ time step by getting the surrogate recommender retrained with the generated fake data up to time $t$ and getting the performance of $Unfair_{F1}$ on the validation dataset:
\begin{equation}\label{Eq:reward}
r_t =\left\{\begin{array}{l}
Unfair_{F1}(t), (t + 1) \% \tau = 0 \\
0, otherwise
\end{array}\right.
\end{equation}
\end{itemize}
Given the above formulations, we aim to learn an item selection policy $\pi^{item}_{\phi}$ with parameter $\phi$ and the gender selection policy $\pi^g_{\psi}$ with parameter $\psi$ to maximize the expected cumulative reward as follows: $   \max_{\pi^{item}_{\phi}, \pi^g_{\psi}} \mathbb{E}_{\hat{x}_{u_i} \sim \pi}\left[\sum_{t=0}^{T - 1} \gamma^t r_t\right],$
where $T$ is the length of the generated attacking trajectory comprising all the fake user profiles.

\begin{figure*}[t]
	\centering 
	\includegraphics[width=0.85\linewidth]{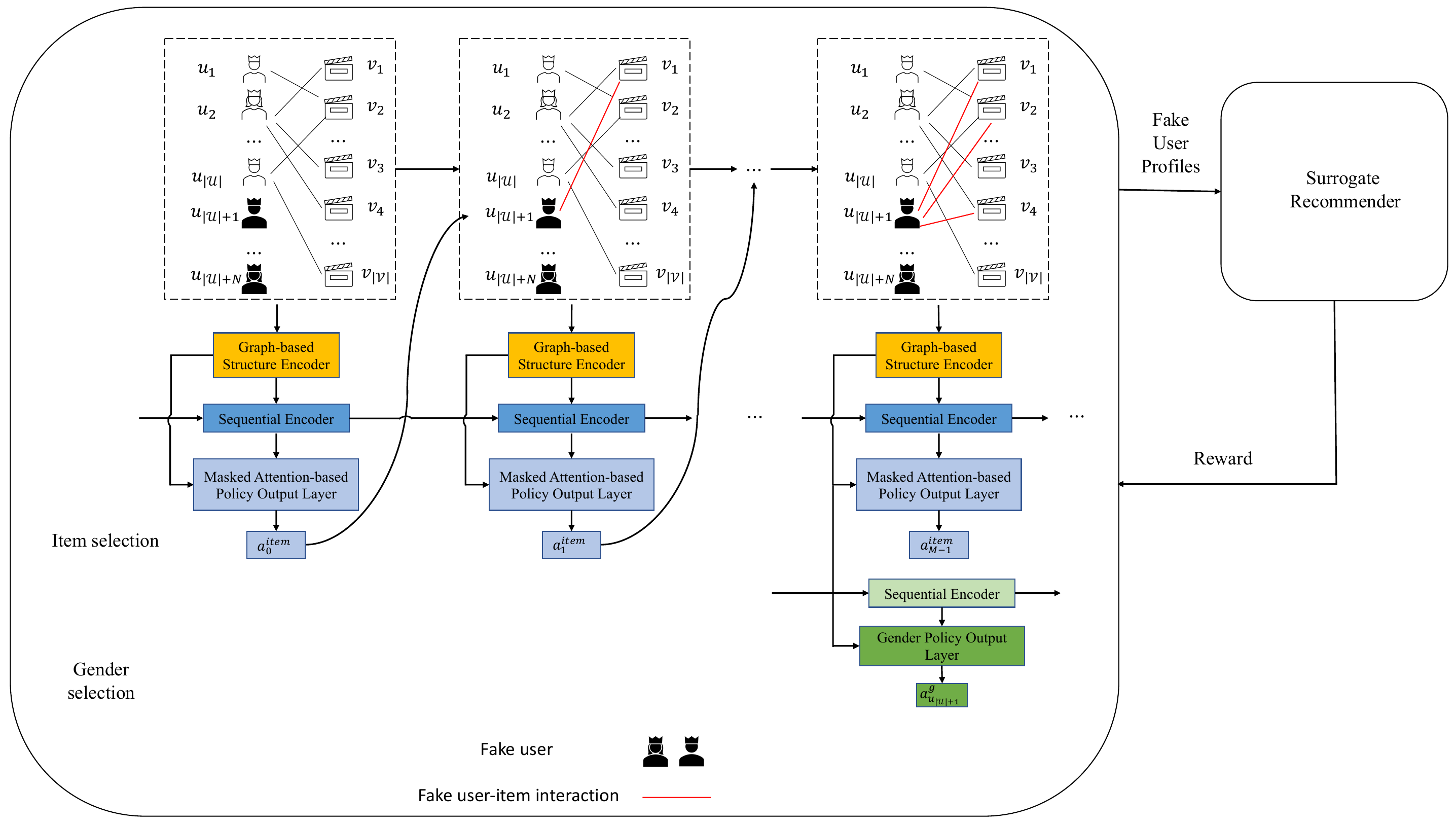}
	\caption{Overview of the proposed framework.}
	\label{fig:overview}
\end{figure*}

\subsection{Overview}
As shown in Figure \ref{fig:overview}, our proposed framework mainly contains five modules: the graph-based structure encoder, the sequential encoder, the masked attention-based policy output layer, gender policy output layer and the surrogate recommender. The user-item interaction data can be naturally seen as a bipartite graph \cite{huang2011does}. The injected fake user-item interaction will influence the original user-item interaction graph by adding the new fake user nodes and new edges to the graph. To capture the structural dependency, we propose to use the graph-based structure encoder to learn the graphical representation with injected fake user profiles. As the fake item is generated sequentially, we utilize a sequential encoder to model the sequential dependency among the generated fake items. Based on the learned state representation with structural and sequential dependency, the masked attention-based policy output layer yields selected item $a^{item}_t$ at time step $t$. As the target recommender system may incorporate the fairness-aware training and leverage the gender attribute of the user profile, we further design a gender policy to determine the gender attribute of the fake user profile. For determining the gender selection action $a^g_{u_i}$ associated with an fake user profile \(u_i\), the gender policy output layer designates a gender at predefined intervals, specifically every \(M\) steps. Within the constraints of a black-box attack scenario, the attacker does not have the knowledge to the target recommender system. To facilitate feedback essential for the training of the policy $\pi^{item}$ and $\pi^{g}$, fake user profiles are injected to a surrogate recommender, which subsequently provides the necessary reward by getting retrained with the injected data. To simulate the potential fairness-aware recommenders and induce a more potent fairness attack, we incorporate the fairness-aware training to the surrogate recommender.
\subsection{Surrogate Recommender}
As shown in the \textit{Attacker's Knowledge}, the target recommender system is a black box to the attacker. A prevalent strategy among adversarial attackers is to construct a local surrogate recommender, wherein the optimization of fake user profiles is executed through attacks on this surrogate. For our study, we utilize one representative recommender in the implicit feedback setting: matrix factorization with bayesian personalized ranking loss (BPR) \cite{rendle2012bpr}. The training loss of BPR with injected fake user profiles is represented as $\mathcal{L}\left(\widetilde{X}, \theta\right)$, where $\widetilde{X} = X \cup \hat{X}$ is the combination of the original training data and the injected fake user profiles. To emulate a more potent attack on the recommender system with possible fairness-aware optimization, we further incorporate a fairness regularization loss into the training objective of the surrogate recommender as follows:

\begin{equation}\label{Eq:fair_train}
\begin{aligned}
    &\mathcal{L}_{\text {fair\_train }} = \mathcal{L} + \lambda \mathcal{L}_{\text {fair }}, \\
    &\mathcal{L}_{\text {fair }}\left(\widetilde{X}, \theta\right)=\left|\mathcal{L}\left(\widetilde{X}_{G_1}, \theta\right)-\mathcal{L}\left(\widetilde{X}_{G_2}, \theta\right)\right|,
\end{aligned}
\end{equation}
where $\widetilde{X}_{G_1}$ and $\widetilde{X}_{G_2}$ are the user profiles among the $\widetilde{X}$ with sensitive attribute $G_1$ and $G_2$ respectively. $\mathcal{L}_{\text {fair }}$ is further smoothed using Huber loss \cite{huber1992robust}. $\lambda$ is the weight of the fairness regularization and is set to 1 in the surrogate recomomender.
\subsection{Graph-based Structure Encoder}
To capture the structural dependency within the poisoned user-item interaction graph, we propose to use a multi-layer graph convolutional network (GCN) \cite{kipf2016semi} to derive the graph representation at time step \( t \). Specifically, the representation of nodes in the \( (l + 1) \)-th layer, denoted as \( H^{(l+1)}_t \), is computed as:
\begin{equation}
    H^{(l+1)}_t = \tilde{D}_t^{-\frac{1}{2}} \tilde{A}_t \tilde{D}_t^{-\frac{1}{2}} H^{(l)}_t W^{(l)},
\end{equation}
where \( \tilde{D}_{t, ii} \) is a diagonal matrix at time $t$ with the element \( \tilde{D}_{t, ii} = \sum_j \tilde{A}_{t, ij} \). The matrix \( \tilde{A}_t = A_t + I_N \) represents the adjacency matrix of the poisoned user-item graph with added self-loops. This poisoned user-item graph integrates the original user-item interaction graph with the generated fake user nodes and user-item interactions up to time step \( t \). Here, \( I_N \) is the identity matrix, \( W^{(l)} \) denotes the parameter matrix for the \( l \)-th layer, and \( H^0_t \in \mathbb{R}^{(|\mathcal{U}| + N + |\mathcal{V}|) \times h} \) is the embedding matrix for all nodes in the poisoned user-item interaction graph with $h$ represents the number of hidden units.

The final graph-based representation is \( H^L_t \), with \( L = 2 \) indicating the number of GCN layers. Leveraging the graph-based structure encoder, the structure-aware representation for the most recently selected fake item \( a_{t - 1}^{item} \) is given by \( z_t^{item} = H_t^L(a_{t - 1}^{item}) \), acquired from the node representation matrix \( H_t^L \). Since fake user profiles are generated sequentially and each has a length of \( M \), the relevant fake user profile at time \( t \) is \( u_{|\mathcal{U}| + \lfloor\frac{t}{M}\rfloor + 1} \). Thus, the structure-aware representation for the fake user node at \( t \) is \( z_t^{user} = H_t^L({|\mathcal{U}| + \lfloor\frac{t}{M}\rfloor + 1}) \), ascertained from \( H_t^L \).

For global structural information, we employ the averaged node embedding across the entire graph, given by:
\begin{equation}
    \Bar{G}_t = \frac{1}{|\mathcal{U}| + N + |\mathcal{V}|} \sum_{i = 1}^{|\mathcal{U}| + N + |\mathcal{V}|} H_t^L(i),
\end{equation}
where $H_t^L(i)$ represents the $i$-th node embedding representation.

\subsection{Sequential Encoder}
As the fake items are generated sequentially, we utilize a recurrent neural network with Long Short Term Memory (LSTM) unit \cite{hochreiter1997long} to model the sequential dependency among the generated fake items. The sequential dependency is computed with the structure-aware representation of the last generated fake item $z_t^{item}$ and the previous hidden state $h_{t - 1}^{item}$. Specifically, the learned hidden state $h_t^{item}$ is computed as follows:
\begin{equation}
\begin{gathered}
i_t=\sigma\left(W_i\left[z_t^{item}, h_{t-1}^{item}\right]+b_i\right), \\
f_t=\sigma\left(W_f\left[z_t^{item}, h_{t-1}^{item}\right]+b_f\right), \\
\tilde{C}_t=\tanh \left(W_C\left[z_t^{item}, h_{t-1}^{item}\right]+b_C\right), \\
C_t=f_t \odot C_{t-1}+i_t \odot \tilde{C}_t \\
o_t=\sigma\left(W_o\left[z_t^{item}, h_{t-1}^{item}\right]+b_o\right), \\
h_t^{item}=o_t \odot \tanh \left(C_t\right),
\end{gathered}
\end{equation}
where $W_i, W_f, W_C, W_o$ are parameter matrices; $b_i, b_f, b_C, b_o$ are parameter of biases. $\odot$ represents element-wise multiplication and $\sigma$ is the sigmoid function. The above update process is denoted in short as:
\begin{equation}
h_t^{item}=L S T M\left(z_t^{item}, h_{t-1}^{item}\right).
\end{equation}
\subsection{Masked Attention-based Output Layer}
Based on the above learned representation with the structure and sequential dependency, the masked attention-based output layer aims to output the next selected fake item. Specifically, we utilize an masked attention mechanism \cite{kool2018attention} to determine the selection weight of different items. The weight for the $j$-th item is calculated as follows:
\begin{equation}
\begin{aligned}
    \omega_{j}&= \begin{cases}C \cdot \tanh \left(\frac{q^T k_j}{\sqrt{d_{\mathrm{k}}}}\right) & \text { if } j \neq a^{item}_{t^{\prime}} \quad \forall t^{\prime} \in \hat{u}(t) \\ -\infty \text { otherwise. }\end{cases}
\end{aligned}
\end{equation}
where $q$ is the query representation, $k_j$ is the key representation of the $j$-th item such that $v_j \in \mathcal{V}$ and $d_k$ is the dimensionality of the query and key representation. To ensure each individual fake user profile selects different fake items, we utilize the mask mechanism to mask out the already selected fake items inside the fake user profile and $\hat{u}({t})$ represents the time indexes that belong to the same fake user profile before time $t$. $C = 10$ is the constant. The query $q$ and key $k_j$ are defined as follows:
\begin{equation}
\begin{aligned}
    q &=W^Q h_{(c)}, \\
    h_{(c)} &= \left[h_t^{item}\|z^{user}_t\|\bar{G}_t\right], \\
    k_i&=W^K H_t^L(i) \quad v_i \in \mathcal{V},
\end{aligned}
\end{equation}
where $\|$ is the concatenation operation. $h_{(c)}$ is the context representation by incorporating the representation of last selected fake item $h_t^{item}$ considering structural and sequential dependency, the current fake user representation $z^{user}_t$ and global representation of the current poisoned user-item interaction graph $\bar{G}_t$. $H_t^L(i)$ is $i$-th item node representation. $W^Q$ and $W^K$ are the parameter matrices for the query and key transformation respectively.

Based on the calculated selection weight for each item, the probability of picking the $i$-th item as the selected fake item at time $t$ is computed using a softmax function as follows:
\begin{equation}
\pi^{item}_{\phi}(a_t^{item} = v_i|s_t) =\frac{e^{\omega_{i}}}{\sum_j e^{\omega_{j}}}.
\end{equation}

\subsection{Gender Selection}
The above process demonstrates how to select the fake item for each fake user profile. In this section, we introduce how to determine the gender action for each fake user profile using the gender selection policy network. As shown in Figure \ref{fig:overview}, the gender selection policy outputs the gender action every $M$ steps, where $M$ is the length of the fake user profile. In other words, the gender policy outputs layer decides the gender action at time $t$, where $(t + 1) \% M = 0$. Similar to the item selection policy network, the gender policy network consists of graph-based structure encoder, the sequential encoder and the gender policy output layer. 

The network structure of the graph-based structure encoder is the same as that in the item selection policy network to learn the structural dependency in the poisoned user-item interaction graph. In the following subsections, we detail the sequential encoder and gender policy output layer in the gender policy network.

\subsubsection{Sequential Encoder for Gender Selection}
As the gender policy network aims to decide the gender action at the user profile level (consisting of $M$ fake items), we get the averaged structure-aware item representation $z_t^{g}$ as the input of the sequential encoder as follows:
\begin{equation}
    z_t^{g} = \frac{1}{M}\sum_{i = t - M + 1}^{t} H_t^L(i).
\end{equation}
Then, the sequential dependency in the fake user profile level is modeled using the recurrent neural network with LSTM cell as follows:
\begin{equation}
h_{cur}^{g}=L S T M\left(z_t^{g}, h_{pre}^{g}\right),
\end{equation}
where $h_{cur}^{g}$ is the learned hidden representation for the current fake user profile. $h_{pre}^{g}$ indicates the hidden presentation for the last fake user profile.
\subsubsection{Gender Policy Output Layer}
Based on the structure-aware and sequence-aware representation in the fake user profile level, the gender policy output layer outputs the selection probability over two types of genders (e.g., male and female) using the softmax function as follows:
\begin{equation}
\pi^g_{\psi}\left(\cdot \mid s_t\right)=\operatorname{Softmax}\left(f\left(h_{cur}^g \| z_t^{user}\right)\right),
\end{equation}
where $\|$ is the concatenation operation and $f$ is a two-layer MultiLayer Perceptron (MLP) with ReLU activation.
\subsection{Training}

\renewcommand{\algorithmicrequire}{\textbf{Input:}}
\renewcommand{\algorithmicensure}{\textbf{Output:}}

\begin{algorithm}[t]
\caption{Fairness Attack Policy Training}
\label{alg:model_training}
\begin{algorithmic}[1]
\Require Item selection policy $\pi_\phi$, gender selection policy $\pi_{\psi}$, surrogate recommender parameters $\theta$, number of fake user profiles $N$, number of selected fake items $M$.
\Ensure Learned item selection policy $\pi_{\phi}$ and gender selection policy $\pi_{\psi}$.
\State Initialize parameters $\phi, \psi, \theta$.
\For{epoch $=1$ to max-epochs}
    \State Reset the initial state $s_0$.
    \For{$t = 0$ to $T - 1$}
        \State Select fake item action $a^{item}_t$ with probability $\pi^{item}_{\phi}(a^{item}_t = v_i \mid s_t)$.
        \State Obtain reward $r_t$ using Eq.~\ref{Eq:reward}.
        \State Update state $s_{t+1} \gets \{s_t, a^{item}_t\}$.
        \If{$(t+1) \bmod M = 0$}
            \State Output gender action $a^g_t$ according to $\pi^g_{\psi}$.
        \EndIf
    \EndFor
    \State Update $\pi^{item}_{\phi}$ and $\pi^g_{\psi}$ using Eq.~\ref{Eq:ppo_loss}.
\EndFor
\end{algorithmic}
\end{algorithm}

Finally, the item selection policy $\pi_\phi$ and gender selection policy $\pi_\psi$ are learned using the widely used Proximal Policy Optimization (PPO) \cite{schulman2017proximal} method. Each of them is optimized using gradient ascent by maximizing the objective $\mathcal{L}^{p p o}(\varphi)$, where $\varphi$ can be $\phi$ or $\psi$:

\begin{equation}\label{Eq:ppo_loss}
\begin{gathered}
    \mathcal{L}^{p p o}(\varphi)=\sum_{t=0}^{T-1}\min (\operatorname{ratio}_t(\varphi) \hat{R}_t, \operatorname{clip}(\operatorname{ratio}_t(\varphi), 1-\epsilon, 1+\epsilon) \hat{R}_t] \\
\operatorname{ratio}_t(\varphi)=\frac{\pi_{\varphi}\left(a_t \mid s_t\right)}{\pi_{\varphi_{\text {old }}}\left(a_t \mid s_t\right)} \\
\hat{R}_t=\sum_{h=t}^{T-1} \gamma^{h-t} r_h,
\end{gathered}
\end{equation}
where $\epsilon$ is a hyperparameter. $\varphi_{\text {old }}$ denotes the old parameters of the policy network and the clip operation restricts ratio $\operatorname{ratio}_t(\varphi)$ to be in the range $[1-\epsilon, 1+\epsilon]$ to avoid drastically policy update. Specifically, the pseudocode of the training algorithm is shown in Algorithm \ref{alg:model_training}.

\section{Experiment}
In this section, we evaluate the effectiveness of our Structure-Aware Reinforcement Learning-Based Fairness Attack (SRLFA) approach using two publicly available datasets and two representative target recommendation models. We aim to address the following \textbf{Research Questions}:
\begin{itemize}
\item \textbf{RQ1:} Does our proposed method exacerbate the unfairness in the recommender system?
\item \textbf{RQ2:} How does our fairness attack method perform when the target recommender system incorporates fairness-aware training?
\item \textbf{RQ3:} What is the performance of the fairness attack method when only a subset of the training data is available?
\item \textbf{RQ4:} How does the number of fake user profiles \(N\) and interactions per fake user profile \(M\) influence the sensitivity of our proposed fairness attack method?
\end{itemize}

\subsection{Datasets}
\begin{table}[t]
\centering
\caption{Statistics of the real-world datasets.}\label{tab:data_stat}
\begin{tabular}{c|cccc}
\hline
Datasets  & \#Users  & \#Items & \#Interactions & Density \\ \hline
MovieLens & 6,040   & 3,706   & 1,000,209      & 4.47\%  \\
Last.fm   & 587     & 7,782 & 160,427        & 3.51\%  \\ \hline
\end{tabular}
\end{table}
We evaluate the proposed SRLFA method on two widely used real-world datasets. Table~\ref{tab:data_stat} presents the statistical details of these datasets. Below is a concise description of both datasets:
\begin{itemize}
    \item \textbf{MovieLens}: This dataset is sourced from the MovieLens-1M user-item interactions in the film domain \cite{harper2015movielens}. Following conventional methodologies \cite{he2017neural, he2020lightgcn}, numeric ratings convert to implicit feedback, indicating user interactions with specific items. The dataset comprises 4,331 male and 1,709 female users.
    
    \item \textbf{Last.fm}: This dataset captures user-song interactions, reflecting user listening preferences up to May 5th, 2009 (\footnote{http://ocelma.net/MusicRecommendationDataset/lastfm-1K.html}). We utilized data from the final month and excluded cold-start users and items with fewer than 10 interactions. Of the 587 users, 319 are male, 220 are female, with the remainder having unspecified genders.
\end{itemize}

For each user in both datasets, interactions are chronologically ordered. The initial 80\% of interactions are used for training, the subsequent 10\% for validation, and the final 10\% for testing.

\begin{table*}[t]
\caption{Fairness attack performance on the NCF recommender in the MovieLens dataset. "None" represents the recommendation results before fairness attack. All the metric values are multiplied by 100. The bold number indicates the best result.}\label{tab:ncf_ml_1m}
\centering
\begin{tabular}{c|ccc|ccc|ccc}
\hline
             & \multicolumn{3}{c|}{NDCG}                                               & \multicolumn{3}{c|}{Precision}                                          & \multicolumn{3}{c}{F1}                                                  \\ \hline
             & \multicolumn{1}{c|}{Male} & \multicolumn{1}{c|}{Female} & Unfair        & \multicolumn{1}{c|}{Male} & \multicolumn{1}{c|}{Female} & Unfair        & \multicolumn{1}{c|}{Male} & \multicolumn{1}{c|}{Female} & Unfair        \\ \hline
None         & \multicolumn{1}{c|}{5.25} & \multicolumn{1}{c|}{5.16}   & 0.09          & \multicolumn{1}{c|}{1.73} & \multicolumn{1}{c|}{1.67}   & 0.06          & \multicolumn{1}{c|}{3.00} & \multicolumn{1}{c|}{2.91}   & 0.08          \\
Revisit      & \multicolumn{1}{c|}{5.53} & \multicolumn{1}{c|}{5.59}   & 0.06          & \multicolumn{1}{c|}{1.81} & \multicolumn{1}{c|}{1.69}   & 0.12          & \multicolumn{1}{c|}{3.13} & \multicolumn{1}{c|}{2.96}   & 0.17          \\
AttackMLFair & \multicolumn{1}{c|}{5.47} & \multicolumn{1}{c|}{5.55}   & 0.08          & \multicolumn{1}{c|}{1.80} & \multicolumn{1}{c|}{1.69}   & 0.11          & \multicolumn{1}{c|}{3.12} & \multicolumn{1}{c|}{2.96}   & 0.16          \\
SRLFA        & \multicolumn{1}{c|}{5.61} & \multicolumn{1}{c|}{5.43}   & \textbf{0.18} & \multicolumn{1}{c|}{1.81} & \multicolumn{1}{c|}{1.66}   & \textbf{0.15} & \multicolumn{1}{c|}{3.14} & \multicolumn{1}{c|}{2.91}   & \textbf{0.23} \\ \hline
\end{tabular}
\end{table*}

\begin{table*}[t]
\caption{Fairness attack performance on the LightGCN recommender in the MovieLens dataset. "None" represents the recommendation results before fairness attack. All the metric values are multiplied by 100. The bold number indicates the best result.}\label{tab:lightgcn_ml_1m}
\centering
\begin{tabular}{c|ccc|ccc|ccc}
\hline
             & \multicolumn{3}{c|}{NDCG}                                               & \multicolumn{3}{c|}{Precision}                                          & \multicolumn{3}{c}{F1}                                                  \\ \hline
             & \multicolumn{1}{c|}{Male} & \multicolumn{1}{c|}{Female} & Unfair        & \multicolumn{1}{c|}{Male} & \multicolumn{1}{c|}{Female} & Unfair        & \multicolumn{1}{c|}{Male} & \multicolumn{1}{c|}{Female} & Unfair        \\ \hline
None         & \multicolumn{1}{c|}{6.06} & \multicolumn{1}{c|}{5.87}   & 0.19          & \multicolumn{1}{c|}{2.02} & \multicolumn{1}{c|}{1.81}   & 0.21          & \multicolumn{1}{c|}{3.48} & \multicolumn{1}{c|}{3.16}   & 0.32          \\
Revisit      & \multicolumn{1}{c|}{6.06} & \multicolumn{1}{c|}{5.86}   & 0.20          & \multicolumn{1}{c|}{2.04} & \multicolumn{1}{c|}{1.82}   & 0.22          & \multicolumn{1}{c|}{3.52} & \multicolumn{1}{c|}{3.18}   & 0.34          \\
AttackMLFair & \multicolumn{1}{c|}{5.97} & \multicolumn{1}{c|}{5.81}   & 0.16          & \multicolumn{1}{c|}{2.02} & \multicolumn{1}{c|}{1.83}   & 0.19          & \multicolumn{1}{c|}{3.48} & \multicolumn{1}{c|}{3.19}   & 0.29          \\
SRLFA        & \multicolumn{1}{c|}{5.90} & \multicolumn{1}{c|}{5.63}   & \textbf{0.27} & \multicolumn{1}{c|}{1.99} & \multicolumn{1}{c|}{1.71}   & \textbf{0.28} & \multicolumn{1}{c|}{3.43} & \multicolumn{1}{c|}{3.00}   & \textbf{0.43} \\ \hline
\end{tabular}
\end{table*}

\begin{table*}[t]
\caption{Fairness attack performance on the NCF recommender in the Last.fm dataset. "None" represents the recommendation results before fairness attack. All the metric values are multiplied by 100. The bold number indicates the best result.}\label{tab:NCF_last_fm}
\centering
\begin{tabular}{c|ccc|ccc|ccc}
\hline
             & \multicolumn{3}{c|}{NDCG}                                                & \multicolumn{3}{c|}{Precision}                                           & \multicolumn{3}{c}{F1}                                                   \\ \hline
             & \multicolumn{1}{c|}{Male}  & \multicolumn{1}{c|}{Female} & Unfair        & \multicolumn{1}{c|}{Male}  & \multicolumn{1}{c|}{Female} & Unfair        & \multicolumn{1}{c|}{Male}  & \multicolumn{1}{c|}{Female} & Unfair        \\ \hline
None         & \multicolumn{1}{c|}{24.64} & \multicolumn{1}{c|}{24.44}  & 0.20          & \multicolumn{1}{c|}{10.85} & \multicolumn{1}{c|}{9.06}   & 1.79          & \multicolumn{1}{c|}{16.06} & \multicolumn{1}{c|}{14.06}  & 2.00          \\
Revisit      & \multicolumn{1}{c|}{24.81} & \multicolumn{1}{c|}{21.77}  & \textbf{3.04}          & \multicolumn{1}{c|}{11.05} & \multicolumn{1}{c|}{8.96}   & 2.09          & \multicolumn{1}{c|}{16.39} & \multicolumn{1}{c|}{13.81}  & 2.58          \\
AttackMLFair & \multicolumn{1}{c|}{24.17} & \multicolumn{1}{c|}{22.58}  & 1.59          & \multicolumn{1}{c|}{10.66} & \multicolumn{1}{c|}{8.99}   & 1.67          & \multicolumn{1}{c|}{15.79} & \multicolumn{1}{c|}{13.94}  & 1.85          \\
SRLFA        & \multicolumn{1}{c|}{23.67} & \multicolumn{1}{c|}{21.46}  & 2.21 & \multicolumn{1}{c|}{10.89} & \multicolumn{1}{c|}{8.75}   & \textbf{2.14} & \multicolumn{1}{c|}{16.17} & \multicolumn{1}{c|}{13.46}  & \textbf{2.71} \\ \hline
\end{tabular}
\end{table*}

\begin{table*}[t]
\caption{Fairness attack performance on the LightGCN recommender in the Last.fm dataset. "None" represents the recommendation results before fairness attack. All the metric values are multiplied by 100. The bold number indicates the best result.}\label{tab:lightgcn_last_fm}
\centering
\begin{tabular}{c|ccc|ccc|ccc}
\hline
             & \multicolumn{3}{c|}{NDCG}                                                & \multicolumn{3}{c|}{Precision}                                           & \multicolumn{3}{c}{F1}                                                   \\ \hline
             & \multicolumn{1}{c|}{Male}  & \multicolumn{1}{c|}{Female} & Unfair        & \multicolumn{1}{c|}{Male}  & \multicolumn{1}{c|}{Female} & Unfair        & \multicolumn{1}{c|}{Male}  & \multicolumn{1}{c|}{Female} & Unfair        \\ \hline
None         & \multicolumn{1}{c|}{26.01} & \multicolumn{1}{c|}{23.86}  & 2.15          & \multicolumn{1}{c|}{10.45} & \multicolumn{1}{c|}{8.70}   & 1.75          & \multicolumn{1}{c|}{15.95} & \multicolumn{1}{c|}{13.63}  & 2.31          \\
Revisit      & \multicolumn{1}{c|}{23.15} & \multicolumn{1}{c|}{20.43}  & 2.72          & \multicolumn{1}{c|}{10.01} & \multicolumn{1}{c|}{8.47}   & 1.54          & \multicolumn{1}{c|}{15.11} & \multicolumn{1}{c|}{13.21}  & 1.90          \\
AttackMLFair & \multicolumn{1}{c|}{22.66} & \multicolumn{1}{c|}{20.78}  & 1.88          & \multicolumn{1}{c|}{9.97}  & \multicolumn{1}{c|}{8.53}   & 1.44          & \multicolumn{1}{c|}{15.01} & \multicolumn{1}{c|}{13.18}  & 1.83          \\
SRLFA        & \multicolumn{1}{c|}{23.29} & \multicolumn{1}{c|}{19.27}  & \textbf{4.02} & \multicolumn{1}{c|}{10.04} & \multicolumn{1}{c|}{8.02}   & \textbf{2.02} & \multicolumn{1}{c|}{15.13} & \multicolumn{1}{c|}{12.52}  & \textbf{2.61} \\ \hline
\end{tabular}
\end{table*}

\subsection{Experimental Setup}
\subsubsection{Evaluation Metrics}
To evaluate the performance of the fairness attack, we utilize the metrics $Unfair_{\mathcal{\text{NDCG}}}$, $Unfair_{\mathcal{\text{Precision}}}$, and $Unfair_{\mathcal{\text{F1}}}$, as defined in Eq. \ref{eq:unfair_metric_define}. These metrics are considered within the context of a recommendation list size of \( K=50 \). Additionally, we present recommendation results separately for male and female groups.
Specifically, NDCG for User \( u \) represented as follows:$NDCG@K(u) = \frac{DCG@K(u)}{IDCG@K(u)},
DCG@K(u) = \sum_{i=1}^K \frac{rel_{u, i}}{\log_2(i+1)},$
where \( rel_{u, i} \) designates the relevance of the item at position \( i \) in the top-\( K \) list recommended to user \( u \). If user \( u \) prefers the \( i \)-th item, then \( rel_{u, i} = 1 \); otherwise, \( rel_{u, i} = 0 \). \( IDCG@K \) represents the ideal DCG spanning all possible recommendation lists of length \( K \), serving as a normalization factor. Precision for a user \( u \) can be calculated using the following formula: $Precision@K(u)= \frac{\left|\mathcal{C}_{u, K} \cap I_u\right|}{K},$
where \( \mathcal{C}_{u, K} \) represents the top-\( K \) items recommended to user \( u \) and \( \mathcal{I}_u \) signifies the set of items preferred by user \( u \). The F1 score is the harmonic mean of precision and recall, defined as follows: 
\begin{equation}
\begin{aligned}
    F1@K(u) &= \frac{2 \times Precision@K(u) \times Recall@K(u)}{Precision@K(u)+Recall@K(u)}, \\
    Recall@K(u) &= \frac{\left|\mathcal{C}_{u, K} \cap \mathcal{I}_u\right|}{\left|\mathcal{I}_u\right|}.
\end{aligned}
\end{equation}

\subsubsection{Target Recommendation Models}\label{sec:targe_rec}
We consider two representative recommendation models as the target recommendation model as follows:
\begin{itemize}
\item \textbf{NCF}\cite{he2017neural}: Neural Collaborative Filtering (NCF) combines the shallow matrix factorization layer with multiple-layer perceptron for the next-item prediction, which can learn both linear and non-linear user-item interaction.
\item \textbf{LightGCN}\cite{he2020lightgcn}: LightGCN is a simplified graph convolution network (GCN) for recommendation task by removing the feature transformation and non-linear activation part from the original GCN \cite{kipf2016semi} model.
\end{itemize}
To simulate the fairness-aware recommendation, we also include the fairness regularization in the training of above two recommenders. Specifically, we add a fairness regularization loss to the original training loss of the recommender as shown in Eq. \ref{Eq:fair_train}. The above two recommenders with fairness regularization are denoted as \textbf{NCF-fair} and \textbf{LightGCN-fair} respectively.
\subsubsection{Implementation Details}
For the surrogate recommendation model, the embedding sizes of both user and item were set to 64. The node embedding size in the graph-based structure encoder was set to 64 and 32 for the MovieLens and Last.fm dataest respectively. The hidden sizes of the graph-based structure encoder, the LSTM layer, the masked attention-based output layer, and the gender policy output layer were all configured to 64 and 32 for the MovieLens and Last.fm dataest respectively. Given the shared representation learned by the graph-based structure encoder, its parameters were utilized commonly among $\pi^{item}_{\phi}$ and $\pi^{g}_{\psi}$. The parameters of both $\pi^{item}_{\phi}$ and $\pi^{g}_{\psi}$ were optimized employing the Adam optimizer \cite{kingma2014adam} with a learning rate set at 0.001. For the attacker’s capability, the number of fake user profiles $N$ and the number of interactions per fake user profile $M$ for all the compared methods were set to $N=180,M=160$ in the MovieLens dataset and $N=60,M=300$ in the Last.fm dataset respectively. Further analysis about the the attacker’s capability is shown in Section \ref{Sec:RQ4}. 

For the target reommendation models, the embedding size was set to 32 for the NCF recommender and we utilized three-layer MLP with hidden size 32 and the regularization weight was set to 0.00001. For the LightGCN model, the embedding size was set to 64 and the regularization weight was set to 0.00001. The two recommenders was optimized using Adam \cite{kingma2014adam} with learning rate of 0.001. 
For the MovieLens dataset, we set \(\lambda\) to 10 for NCF-fair and 5 for LightGCN-fair. In the case of the Last.fm dataset, \(\lambda\) was configured to 1 for NCF-fair and 5 for LightGCN-fair.

\begin{table*}[t]
\caption{Fairness attack performance on the NCF-fair recommender in the MovieLens dataset. "None-Fair" represents the fairness-aware recommendation results before fairness attack. All the metric values are multiplied by 100. The bold number indicates the best result.}\label{tab:ncf_fair_ml_1m}
\centering
\begin{tabular}{c|ccc|ccc|ccc}
\hline
             & \multicolumn{3}{c|}{NDCG}                                               & \multicolumn{3}{c|}{Precision}                                          & \multicolumn{3}{c}{F1}                                                  \\ \hline
             & \multicolumn{1}{c|}{Male} & \multicolumn{1}{c|}{Female} & Unfair        & \multicolumn{1}{c|}{Male} & \multicolumn{1}{c|}{Female} & Unfair        & \multicolumn{1}{c|}{Male} & \multicolumn{1}{c|}{Female} & Unfair        \\ \hline
None-Fair    & \multicolumn{1}{c|}{4.42} & \multicolumn{1}{c|}{4.47}   & 0.05          & \multicolumn{1}{c|}{1.33} & \multicolumn{1}{c|}{1.27}   & 0.06          & \multicolumn{1}{c|}{2.36} & \multicolumn{1}{c|}{2.27}   & 0.08          \\
Revisit      & \multicolumn{1}{c|}{4.68} & \multicolumn{1}{c|}{4.66}   & 0.02          & \multicolumn{1}{c|}{1.42} & \multicolumn{1}{c|}{1.31}   & 0.11          & \multicolumn{1}{c|}{2.51} & \multicolumn{1}{c|}{2.33}   & 0.18          \\
AttackMLFair & \multicolumn{1}{c|}{4.62} & \multicolumn{1}{c|}{4.53}   & \textbf{0.09}          & \multicolumn{1}{c|}{1.35} & \multicolumn{1}{c|}{1.28}   & 0.07          & \multicolumn{1}{c|}{2.40} & \multicolumn{1}{c|}{2.28}   & 0.11          \\
SRLFA        & \multicolumn{1}{c|}{4.68} & \multicolumn{1}{c|}{4.76}   & 0.08 & \multicolumn{1}{c|}{1.38} & \multicolumn{1}{c|}{1.25}   & \textbf{0.13} & \multicolumn{1}{c|}{2.44} & \multicolumn{1}{c|}{2.25}   & \textbf{0.20} \\ \hline
\end{tabular}
\end{table*}

\begin{table*}[t]
\caption{Fairness attack performance on the LightGCN-fair recommender in the MovieLens dataset. "None-Fair" represents the fairness-aware recommendation results before fairness attack. All the metric values are multiplied by 100. The bold number indicates the best result.}\label{tab:lightgcn_fair_ml_1m}
\centering
\begin{tabular}{c|ccc|ccc|ccc}
\hline
             & \multicolumn{3}{c|}{NDCG}                                               & \multicolumn{3}{c|}{Precision}                                          & \multicolumn{3}{c}{F1}                                                  \\ \hline
             & \multicolumn{1}{c|}{Male} & \multicolumn{1}{c|}{Female} & Unfair        & \multicolumn{1}{c|}{Male} & \multicolumn{1}{c|}{Female} & Unfair        & \multicolumn{1}{c|}{Male} & \multicolumn{1}{c|}{Female} & Unfair        \\ \hline
None-Fair    & \multicolumn{1}{c|}{6.17} & \multicolumn{1}{c|}{6.08}   & 0.09          & \multicolumn{1}{c|}{1.95} & \multicolumn{1}{c|}{1.75}   & 0.20          & \multicolumn{1}{c|}{3.41} & \multicolumn{1}{c|}{3.11}   & 0.30          \\
Revisit      & \multicolumn{1}{c|}{6.10} & \multicolumn{1}{c|}{6.04}   & 0.06          & \multicolumn{1}{c|}{1.94} & \multicolumn{1}{c|}{1.73}   & 0.21          & \multicolumn{1}{c|}{3.39} & \multicolumn{1}{c|}{3.07}   & 0.32          \\
AttackMLFair & \multicolumn{1}{c|}{6.14} & \multicolumn{1}{c|}{6.11}   & 0.03          & \multicolumn{1}{c|}{1.93} & \multicolumn{1}{c|}{1.76}   & 0.17          & \multicolumn{1}{c|}{3.38} & \multicolumn{1}{c|}{3.12}   & 0.25          \\
SRLFA        & \multicolumn{1}{c|}{6.09} & \multicolumn{1}{c|}{5.95}   & \textbf{0.14} & \multicolumn{1}{c|}{1.95} & \multicolumn{1}{c|}{1.72}   & \textbf{0.23} & \multicolumn{1}{c|}{3.41} & \multicolumn{1}{c|}{3.05}   & \textbf{0.36} \\ \hline
\end{tabular}
\end{table*}

\begin{table*}[t]
\caption{Fairness attack performance on the NCF-fair recommender in the Last.fm dataset. "None-Fair" represents the fairness-aware recommendation results before fairness attack. All the metric values are multiplied by 100. The bold number indicates the best result.}\label{tab:NCF_fair_last_fm}
\centering
\begin{tabular}{c|ccc|ccc|ccc}
\hline
             & \multicolumn{3}{c|}{NDCG}                                                & \multicolumn{3}{c|}{Precision}                                           & \multicolumn{3}{c}{F1}                                                   \\ \hline
             & \multicolumn{1}{c|}{Male}  & \multicolumn{1}{c|}{Female} & Unfair        & \multicolumn{1}{c|}{Male}  & \multicolumn{1}{c|}{Female} & Unfair        & \multicolumn{1}{c|}{Male}  & \multicolumn{1}{c|}{Female} & Unfair        \\ \hline
None-Fair    & \multicolumn{1}{c|}{26.08} & \multicolumn{1}{c|}{26.13}  & 0.05          & \multicolumn{1}{c|}{11.10} & \multicolumn{1}{c|}{9.38}   & 1.72          & \multicolumn{1}{c|}{16.44} & \multicolumn{1}{c|}{14.80}  & 1.64          \\
Revisit      & \multicolumn{1}{c|}{25.61} & \multicolumn{1}{c|}{25.71}  & 0.10          & \multicolumn{1}{c|}{11.07} & \multicolumn{1}{c|}{9.45}   & 1.62          & \multicolumn{1}{c|}{16.45} & \multicolumn{1}{c|}{14.84}  & 1.61          \\
AttackMLFair & \multicolumn{1}{c|}{25.31} & \multicolumn{1}{c|}{25.96}  & 0.65          & \multicolumn{1}{c|}{10.81} & \multicolumn{1}{c|}{9.59}   & 1.22          & \multicolumn{1}{c|}{16.12} & \multicolumn{1}{c|}{15.13}  & 0.99          \\
SRLFA        & \multicolumn{1}{c|}{26.69} & \multicolumn{1}{c|}{25.44}  & \textbf{1.25} & \multicolumn{1}{c|}{11.24} & \multicolumn{1}{c|}{9.42}   & \textbf{1.82} & \multicolumn{1}{c|}{16.72} & \multicolumn{1}{c|}{14.75}  & \textbf{1.96} \\ \hline
\end{tabular}
\end{table*}

\begin{table*}[t]
\caption{Fairness attack performance on the LightGCN-fair recommender in the Last.fm dataset. "None-Fair" represents the fairness-aware recommendation results before fairness attack. All the metric values are multiplied by 100. The bold number indicates the best result.}\label{tab:lightgcn_fair_last_fm}
\centering
\begin{tabular}{c|ccc|ccc|ccc}
\hline
             & \multicolumn{3}{c|}{NDCG}                                                & \multicolumn{3}{c|}{Precision}                                          & \multicolumn{3}{c}{F1}                                                   \\ \hline
             & \multicolumn{1}{c|}{Male}  & \multicolumn{1}{c|}{Female} & Unfair        & \multicolumn{1}{c|}{Male} & \multicolumn{1}{c|}{Female} & Unfair        & \multicolumn{1}{c|}{Male}  & \multicolumn{1}{c|}{Female} & Unfair        \\ \hline
None-Fair    & \multicolumn{1}{c|}{23.74} & \multicolumn{1}{c|}{23.87}  & 0.13          & \multicolumn{1}{c|}{9.94} & \multicolumn{1}{c|}{9.03}   & 0.91          & \multicolumn{1}{c|}{15.02} & \multicolumn{1}{c|}{14.09}  & 0.93          \\
Revisit      & \multicolumn{1}{c|}{22.07} & \multicolumn{1}{c|}{22.70}  & 0.63          & \multicolumn{1}{c|}{9.54} & \multicolumn{1}{c|}{9.12}   & 0.42          & \multicolumn{1}{c|}{14.34} & \multicolumn{1}{c|}{14.14}  & 0.20          \\
AttackMLFair & \multicolumn{1}{c|}{22.09} & \multicolumn{1}{c|}{21.95}  & 0.14          & \multicolumn{1}{c|}{9.57} & \multicolumn{1}{c|}{8.88}   & 0.69          & \multicolumn{1}{c|}{14.27} & \multicolumn{1}{c|}{13.74}  & 0.54          \\
SRLFA        & \multicolumn{1}{c|}{21.08} & \multicolumn{1}{c|}{17.67}  & \textbf{3.41} & \multicolumn{1}{c|}{9.35} & \multicolumn{1}{c|}{7.61}   & \textbf{1.74} & \multicolumn{1}{c|}{14.06} & \multicolumn{1}{c|}{11.57}  & \textbf{2.49} \\ \hline
\end{tabular}
\end{table*}

\subsubsection{Baseline Methods}
As there are no existing works studying the fairness attacks on recommender systems. We adapt two representative attacking methods for the fairness attack in recommender systems:
\begin{itemize}
    \item \textbf{Revisit} \cite{tang2020revisiting}:
    This is one state-of-the-art method for attacking the performance of recommendation model by promoting the target items. We modify the attack objective to decrease the fairness of the surrogate recommender (i.e., weighted matrix factorization model). For fair comparison, the training loss is the same as $\mathcal{L}_{\text {fair\_train }}$ used in our model.
    \item \textbf{AttackMLFair} \cite{mehrabi2021exacerbating}: This is one state-of-the-art method for attacking the fairness of machine learning classification algorithms. This method modifies malicious data points to reduce fairness through an influence-based attack strategy \cite{koh2022stronger}. We recalibrate the training loss in the surrogate recommender (i.e., weighted matrix factorization model) to match our adopted training loss in Eq. \ref{Eq:fair_train}. Furthermore, we alter the influence estimation, focusing on compromising the fairness of the recommender rather than merely the classification fairness.
\end{itemize}

\begin{figure*}[ht]
	\centering 
	\includegraphics[width=0.5\linewidth]{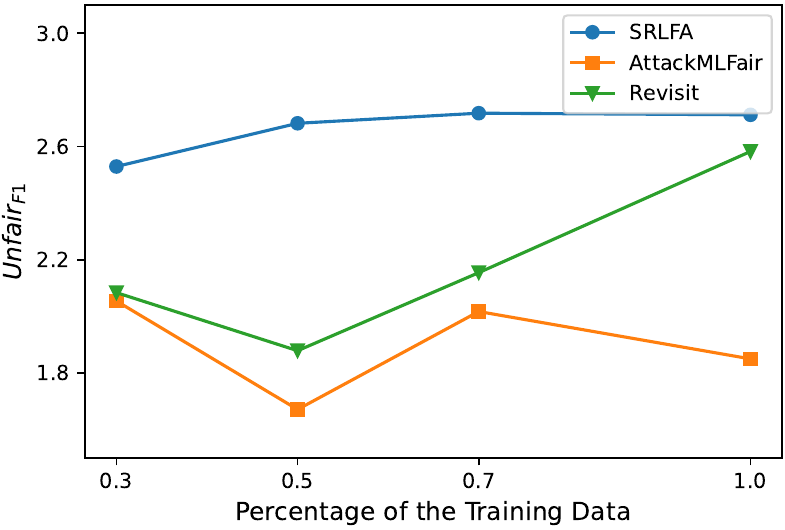}
	\caption{Impact of the percentage of training data on attacking NCF on Last.fm dataset.}
	\label{fig:train_data_ratio}
\end{figure*}

\subsection{Fairness Attack Results on Recommender System (RQ1)}\label{sec:rq1}
To answer RQ1, we compare the proposed method with the baseline methods on two target recommendation models over two real-world datasets. Table \ref{tab:ncf_ml_1m} - \ref{tab:lightgcn_last_fm} show the comparison results. In addition to unfairness metrics, the tables also show the average recommendation performance for both male and female users. We have the following observations. First, our proposed method can exacerbate the unfairness of all the target recommenders on the three unfairness metrics across two datasets. For instance, as highlighted in Table \ref{tab:ncf_ml_1m}, our proposed SRLFA exacerbates the unfairness of NCF by 150.00\% based on the $Unfair_{\text{precision}}$ metric. On average, our proposed fairness attack method amplifies unfairness by averages of 297.24\% on $Unfair_{\text{NDCG}}$, 53.72\% on $Unfair_{\text{precision}}$ and 67.48\% on $Unfair_{\text{F1}}$ across the two target recommenders and two datasets. Second, our method outperforms all the baseline methods on all the unfairness metrics across all the target recommender and dataset combinations except the $Unfair_{\text{NDCG}}$ in Table \ref{tab:NCF_last_fm}. Aggregated results indicate that SRLFA outperforms the best baseline by margins of 36.21\%, 20.48\%, and 25.91\% for $Unfair_{\text{NDCG}}$, $Unfair_{\text{precision}}$, and $Unfair_{\text{F1}}$ respectively. Third, In contrast to baseline methods, our method appears to further diminish recommendation performance for the disadvantaged group (specifically, the female group, as evidenced in the "None" results). This deepens the performance disparity between male and female users. For instance, in Table \ref{tab:lightgcn_ml_1m}, our approach reduces the F1 metric performance for females from 3.16 to 3.00, marking the most pronounced decline in female user performance in comparison to baseline methods. These results clearly demonstrate that our fairness attack method can significantly exacerbate the unfairness of recommender system.

\subsection{Fairness Attack Results on Fairness-aware Recommender System (RQ2)}
The above attack results demonstrate that the fairness attacks can indeed harm the fairness of recommender system. An intriguing question is how does the proposed fairness attack method perform when the target recommender system incorporate fairness aware training? (RQ2) To explore this, we assess the impact of fairness attacks on two fairness-aware recommenders: NCF-fair and LightGCN-fair, as outlined in Section \ref{sec:targe_rec}. The ensuing fairness attack outcomes are detailed in Tables \ref{tab:ncf_fair_ml_1m}-\ref{tab:lightgcn_fair_last_fm}. From these tables, we have the following observations. First, fairness-aware recommenders appear to attenuate the efficacy of fairness attacks when compared to their non-fairness-trained counterparts. For instance, LightGCN-fair reduced the $Unfair_{\text{NDCG}}$ induced by our SRLFA method from 4.02 (Table \ref{tab:lightgcn_last_fm}) to 3.41 (Table \ref{tab:lightgcn_fair_last_fm}). Nonetheless, SRLFA remains capable of intensifying the unfairness within these fairness-aware target recommenders. On average, our SRLFA method can exacerbate the unfairness of fairness-aware recommender (e.g., None-Fair) by 1259.66\%, 57.17\% and 89.31\% on $Unfair_{\text{NDCG}}$, $Unfair_{\text{precision}}$ and $Unfair_{\text{F1}}$ respectively. Second, the SRLFA method consistently surpasses all baseline methods across most unfairness metrics, with exceptions like $Unfair_{\text{NDCG}}$ in Table \ref{tab:ncf_fair_ml_1m}. On average, it exceeds the performance of the best baseline attack methods by 163.95\% on $Unfair_{\text{NDCG}}$, 48.06\% on $Unfair_{\text{precision}}$, and 101.62\% on $Unfair_{\text{F1}}$ across two target recommenders and two datasets. Third, mirroring the observations on non-fairness-trained recommenders in Section \ref{sec:rq1}, SRLFA further diminishes recommendation accuracy for disadvantaged groups (e.g., female users), thereby amplifying disparity in recommendation outcomes between male and female user groups.  These results clearly demonstrate that our fairness attack method can still significantly harm the fairness of fairness-aware recommenders.

In the subsequent sections, we provide further analysis of the proposed method and use the fairness attack performance on NCF recommender on Last.fm dataset as an representative example.

\subsection{Impact of the Percentage of Training Data (RQ3)}

In this section, we examine the performance of the proposed fairness attack method when the attacker has access to only a portion of the training data (RQ3). For each user's interactions in the training data, we randomly sample a subset of the interactions to represent the training data observed by the attacker. Fig.\ref{fig:train_data_ratio} illustrates the performance of the $Unfair_{\text{F1}}$ fairness attack on NCF recommender on the Last.fm dataset with different percentages of training data available. As can be seen from Fig. \ref{fig:train_data_ratio}, our proposed SRLFA method surpasses the two baseline attack methods across various percentages of training data. Generally, the performance of the fairness attack improves with an increase in the percentage of available training data. Remarkably, the SRLFA method already achieves relatively strong fairness attack performance with just 30\% of the training data. This outcome underscores the vulnerability of recommender systems' fairness, even when only partial data is accessible to the attacker.

\subsection{Impact of Attacker's Capability of $N$ and $M$ (RQ4)}\label{Sec:RQ4}
\begin{figure*}[ht]
    \centering
    \begin{minipage}[t]{0.46\textwidth}
        \centering
        \includegraphics[width=\linewidth]{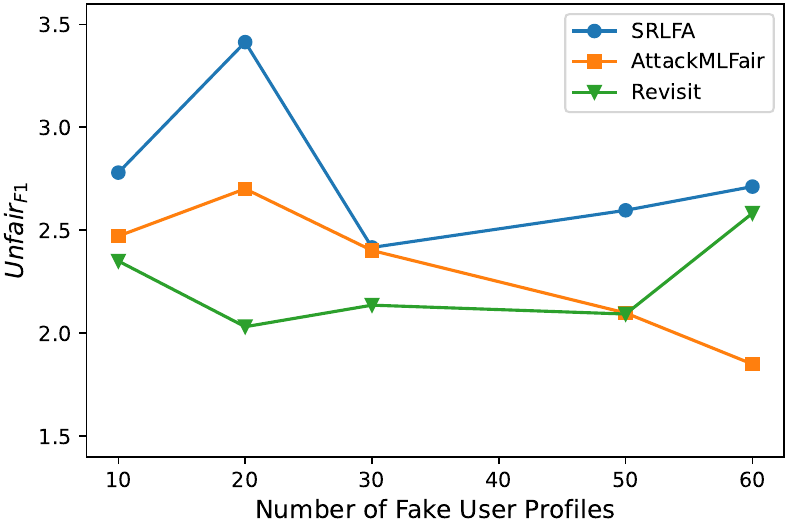}
        \\[1ex]
        \small (a) Impact of $N$
    \end{minipage}
    \hfill
    \begin{minipage}[t]{0.46\textwidth}
        \centering
        \includegraphics[width=\linewidth]{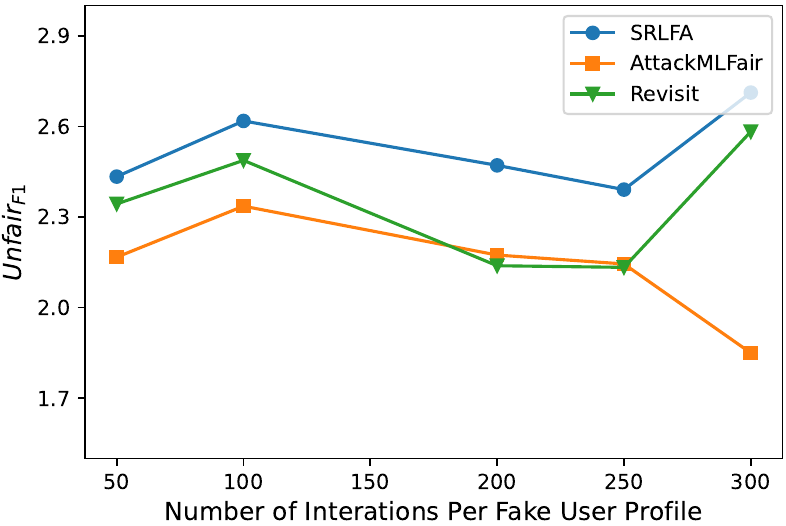}
        \\[1ex]
        \small (b) Impact of $M$
    \end{minipage}

    \caption{Impact of attacker capability $N$ and $M$ on fairness attack performance for the NCF recommender on the Last.fm dataset.}
    \label{fig:impact_n_m}
\end{figure*}

In this section, we examine the influence of the attacker's capability (i.e., the number of generated fake user profiles $N$ and the number of interactions per fake user profile $M$) on the fairness attack performance (RQ4). To address RQ4, we vary both the number of fake user profiles and the number of interactions per fake user; the results of this variation on the fairness attack are illustrated in Fig. \ref{fig:impact_n_m}. As depicted in Fig. \ref{fig:impact_n_m} (a), our proposed SRLFA method outperforms the two baseline attack methods across a range of fake user profiles. The optimal fairness attack performance of SRLFA is attained when the number of fake user profiles is medium (i.e., $N=20$). Likewise, Fig. \ref{fig:impact_n_m} (b) reveals that our fairness attack method continues to surpass the two baseline attack methods across different numbers of interactions per fake user profile, with the peak fairness attack performance achieved when $M = 300$. This value approximates the average number of user interactions in the Last.fm dataset.

\section{Conclusion}
In this paper, we investigate the under-explored problem of fairness attacks on recommender systems and propose a novel Structure-aware Reinforcement Learning based Fairness Attack (SRLFA) method designed to exacerbate the unfairness of recommender systems. Our method models the structural dependency between fake user-item interactions and original user-item interactions. Specifically, we employ a graph-based structure encoder to model this structural dependency. Since fake items are determined sequentially, we then use a recurrent neural network to model the sequential dependency among these fake items. Based on the learned structure-aware and sequence-aware representations of the fake user and item, the item selection policy employs a masked attention mechanism to decide the next fake item. Additionally, as the target recommendation model may employ fairness-aware training and leverage gender attribute information, we further design a gender selection policy to decide the gender information of the generated fake user profile. Given the practical black-box attack setting, we utilize a fairness-aware surrogate model to simulate the potential fairness-aware target recommender and provide feedback for the optimization of the fake user profile. Both the item selection and gender selection policies are jointly trained using reinforcement learning in our proposed method. Extensive experiments on four representative recommender models and two real-world datasets demonstrate that SRLFA can significantly exacerbate system-level unfairness, highlighting critical fairness vulnerabilities and underscoring the importance of red-teaming approaches for proactive fairness risk management in recommender systems.

\bibliographystyle{IEEEtran}
\bibliography{ref}

@article{wang2023survey,
  title={A survey on the fairness of recommender systems},
  author={Wang, Yifan and Ma, Weizhi and Zhang, Min and Liu, Yiqun and Ma, Shaoping},
  journal={ACM Transactions on Information Systems},
  volume={41},
  number={3},
  pages={1--43},
  year={2023},
  publisher={ACM New York, NY}
}

@inproceedings{tang2020revisiting,
  title={Revisiting adversarially learned injection attacks against recommender systems},
  author={Tang, Jiaxi and Wen, Hongyi and Wang, Ke},
  booktitle={Proceedings of the 14th ACM Conference on Recommender Systems},
  pages={318--327},
  year={2020}
}

@inproceedings{zhang2020practical,
  title={Practical data poisoning attack against next-item recommendation},
  author={Zhang, Hengtong and Li, Yaliang and Ding, Bolin and Gao, Jing},
  booktitle={Proceedings of The Web Conference 2020},
  pages={2458--2464},
  year={2020}
}

@book{sutton2018reinforcement,
  title={Reinforcement learning: An introduction},
  author={Sutton, Richard S and Barto, Andrew G},
  year={2018},
  publisher={MIT press}
}

@article{silver2016mastering,
  title={Mastering the game of Go with deep neural networks and tree search},
  author={Silver, David and Huang, Aja and Maddison, Chris J and Guez, Arthur and Sifre, Laurent and Van Den Driessche, George and Schrittwieser, Julian and Antonoglou, Ioannis and Panneershelvam, Veda and Lanctot, Marc and others},
  journal={nature},
  volume={529},
  number={7587},
  pages={484--489},
  year={2016},
  publisher={Nature Publishing Group}
}

@article{rendle2012bpr,
  title={BPR: Bayesian personalized ranking from implicit feedback},
  author={Rendle, Steffen and Freudenthaler, Christoph and Gantner, Zeno and Schmidt-Thieme, Lars},
  journal={arXiv preprint arXiv:1205.2618},
  year={2012}
}

@incollection{huber1992robust,
  title={Robust estimation of a location parameter},
  author={Huber, Peter J},
  booktitle={Breakthroughs in statistics: Methodology and distribution},
  pages={492--518},
  year={1992},
  publisher={Springer}
}

@article{hochreiter1997long,
  title={Long short-term memory},
  author={Hochreiter, Sepp and Schmidhuber, J{\"u}rgen},
  journal={Neural computation},
  volume={9},
  number={8},
  pages={1735--1780},
  year={1997},
  publisher={MIT press}
}

@article{kool2018attention,
  title={Attention, learn to solve routing problems!},
  author={Kool, Wouter and Van Hoof, Herke and Welling, Max},
  journal={arXiv preprint arXiv:1803.08475},
  year={2018}
}

@article{schulman2017proximal,
  title={Proximal policy optimization algorithms},
  author={Schulman, John and Wolski, Filip and Dhariwal, Prafulla and Radford, Alec and Klimov, Oleg},
  journal={arXiv preprint arXiv:1707.06347},
  year={2017}
}

@inproceedings{he2017neural,
  title={Neural collaborative filtering},
  author={He, Xiangnan and Liao, Lizi and Zhang, Hanwang and Nie, Liqiang and Hu, Xia and Chua, Tat-Seng},
  booktitle={Proceedings of the 26th international conference on world wide web},
  pages={173--182},
  year={2017}
}

@inproceedings{he2020lightgcn,
  title={Lightgcn: Simplifying and powering graph convolution network for recommendation},
  author={He, Xiangnan and Deng, Kuan and Wang, Xiang and Li, Yan and Zhang, Yongdong and Wang, Meng},
  booktitle={Proceedings of the 43rd International ACM SIGIR conference on research and development in Information Retrieval},
  pages={639--648},
  year={2020}
}

@article{harper2015movielens,
  title={The movielens datasets: History and context},
  author={Harper, F Maxwell and Konstan, Joseph A},
  journal={Acm transactions on interactive intelligent systems (tiis)},
  volume={5},
  number={4},
  pages={1--19},
  year={2015},
  publisher={Acm New York, NY, USA}
}

@article{kingma2014adam,
  title={Adam: A method for stochastic optimization},
  author={Kingma, Diederik P and Ba, Jimmy},
  journal={arXiv preprint arXiv:1412.6980},
  year={2014}
}

@article{kipf2016semi,
  title={Semi-supervised classification with graph convolutional networks},
  author={Kipf, Thomas N and Welling, Max},
  journal={arXiv preprint arXiv:1609.02907},
  year={2016}
}

@inproceedings{mehrabi2021exacerbating,
  title={Exacerbating algorithmic bias through fairness attacks},
  author={Mehrabi, Ninareh and Naveed, Muhammad and Morstatter, Fred and Galstyan, Aram},
  booktitle={Proceedings of the AAAI Conference on Artificial Intelligence},
  volume={35},
  number={10},
  pages={8930--8938},
  year={2021}
}

@article{koh2022stronger,
  title={Stronger data poisoning attacks break data sanitization defenses},
  author={Koh, Pang Wei and Steinhardt, Jacob and Liang, Percy},
  journal={Machine Learning},
  pages={1--47},
  year={2022},
  publisher={Springer}
}

@inproceedings{yang2017fake,
  title={Fake Co-visitation Injection Attacks to Recommender Systems.},
  author={Yang, Guolei and Gong, Neil Zhenqiang and Cai, Ying},
  booktitle={NDSS},
  year={2017}
}

@article{li2016data,
  title={Data poisoning attacks on factorization-based collaborative filtering},
  author={Li, Bo and Wang, Yining and Singh, Aarti and Vorobeychik, Yevgeniy},
  journal={Advances in neural information processing systems},
  volume={29},
  year={2016}
}

@inproceedings{christakopoulou2019adversarial,
  title={Adversarial attacks on an oblivious recommender},
  author={Christakopoulou, Konstantina and Banerjee, Arindam},
  booktitle={Proceedings of the 13th ACM Conference on Recommender Systems},
  pages={322--330},
  year={2019}
}

@inproceedings{fang2020influence,
  title={Influence function based data poisoning attacks to top-n recommender systems},
  author={Fang, Minghong and Gong, Neil Zhenqiang and Liu, Jia},
  booktitle={Proceedings of The Web Conference 2020},
  pages={3019--3025},
  year={2020}
}

@inproceedings{fang2018poisoning,
  title={Poisoning attacks to graph-based recommender systems},
  author={Fang, Minghong and Yang, Guolei and Gong, Neil Zhenqiang and Liu, Jia},
  booktitle={Proceedings of the 34th annual computer security applications conference},
  pages={381--392},
  year={2018}
}

@inproceedings{song2020poisonrec,
  title={Poisonrec: an adaptive data poisoning framework for attacking black-box recommender systems},
  author={Song, Junshuai and Li, Zhao and Hu, Zehong and Wu, Yucheng and Li, Zhenpeng and Li, Jian and Gao, Jun},
  booktitle={2020 IEEE 36th International Conference on Data Engineering (ICDE)},
  pages={157--168},
  year={2020},
  organization={IEEE}
}

@inproceedings{chirita2005preventing,
  title={Preventing shilling attacks in online recommender systems},
  author={Chirita, Paul-Alexandru and Nejdl, Wolfgang and Zamfir, Cristian},
  booktitle={Proceedings of the 7th annual ACM international workshop on Web information and data management},
  pages={67--74},
  year={2005}
}

@inproceedings{lin2020attacking,
  title={Attacking recommender systems with augmented user profiles},
  author={Lin, Chen and Chen, Si and Li, Hui and Xiao, Yanghua and Li, Lianyun and Yang, Qian},
  booktitle={Proceedings of the 29th ACM international conference on information \& knowledge management},
  pages={855--864},
  year={2020}
}

@inproceedings{solans2020poisoning,
  title={Poisoning attacks on algorithmic fairness},
  author={Solans, David and Biggio, Battista and Castillo, Carlos},
  booktitle={Joint European Conference on Machine Learning and Knowledge Discovery in Databases},
  pages={162--177},
  year={2020},
  organization={Springer}
}

@inproceedings{koh2017understanding,
  title={Understanding black-box predictions via influence functions},
  author={Koh, Pang Wei and Liang, Percy},
  booktitle={International conference on machine learning},
  pages={1885--1894},
  year={2017},
  organization={PMLR}
}

@inproceedings{covington2016deep,
  title={Deep neural networks for youtube recommendations},
  author={Covington, Paul and Adams, Jay and Sargin, Emre},
  booktitle={Proceedings of the 10th ACM conference on recommender systems},
  pages={191--198},
  year={2016}
}

@article{urdaneta2021recommendation,
  title={Recommendation systems for education: Systematic review},
  author={Urdaneta-Ponte, Mar{\'\i}a Cora and Mendez-Zorrilla, Amaia and Oleagordia-Ruiz, Ibon},
  journal={Electronics},
  volume={10},
  number={14},
  pages={1611},
  year={2021},
  publisher={MDPI}
}

@inproceedings{kenthapadi2017personalized,
  title={Personalized job recommendation system at linkedin: Practical challenges and lessons learned},
  author={Kenthapadi, Krishnaram and Le, Benjamin and Venkataraman, Ganesh},
  booktitle={Proceedings of the eleventh ACM conference on recommender systems},
  pages={346--347},
  year={2017}
}

@article{yao2017beyond,
  title={Beyond parity: Fairness objectives for collaborative filtering},
  author={Yao, Sirui and Huang, Bert},
  journal={Advances in neural information processing systems},
  volume={30},
  year={2017}
}

@article{kochling2020discriminated,
  title={Discriminated by an algorithm: a systematic review of discrimination and fairness by algorithmic decision-making in the context of HR recruitment and HR development},
  author={K{\"o}chling, Alina and Wehner, Marius Claus},
  journal={Business Research},
  volume={13},
  number={3},
  pages={795--848},
  year={2020},
  publisher={Springer}
}

@article{spinks2019contemporary,
  title={Contemporary housing discrimination: Facebook, targeted advertising, and the fair housing act},
  author={Spinks, Chandler Nicholle},
  journal={Hous. L. Rev.},
  volume={57},
  pages={925},
  year={2019},
  publisher={HeinOnline}
}

@article{tobin2019hud,
  title={HUD sues Facebook over housing discrimination and says the company’s algorithms have made the problem worse},
  author={Tobin, Ariana},
  journal={ProPublica (March 28, 2019). Available at https://www. propublica. org/article/hud-sues-facebook-housing-discrimination-advertising-algorithms (last accessed April 29, 2019)},
  year={2019}
}

@inproceedings{beutel2019fairness,
  title={Fairness in recommendation ranking through pairwise comparisons},
  author={Beutel, Alex and Chen, Jilin and Doshi, Tulsee and Qian, Hai and Wei, Li and Wu, Yi and Heldt, Lukasz and Zhao, Zhe and Hong, Lichan and Chi, Ed H and others},
  booktitle={Proceedings of the 25th ACM SIGKDD international conference on knowledge discovery \& data mining},
  pages={2212--2220},
  year={2019}
}

@inproceedings{bose2019compositional,
  title={Compositional fairness constraints for graph embeddings},
  author={Bose, Avishek and Hamilton, William},
  booktitle={International Conference on Machine Learning},
  pages={715--724},
  year={2019},
  organization={PMLR}
}

@inproceedings{ekstrand2018all,
  title={All the cool kids, how do they fit in?: Popularity and demographic biases in recommender evaluation and effectiveness},
  author={Ekstrand, Michael D and Tian, Mucun and Azpiazu, Ion Madrazo and Ekstrand, Jennifer D and Anuyah, Oghenemaro and McNeill, David and Pera, Maria Soledad},
  booktitle={Conference on fairness, accountability and transparency},
  pages={172--186},
  year={2018},
  organization={PMLR}
}

@inproceedings{wu2021triple,
  title={Triple adversarial learning for influence based poisoning attack in recommender systems},
  author={Wu, Chenwang and Lian, Defu and Ge, Yong and Zhu, Zhihao and Chen, Enhong},
  booktitle={Proceedings of the 27th ACM SIGKDD Conference on Knowledge Discovery \& Data Mining},
  pages={1830--1840},
  year={2021}
}

@inproceedings{zhang2021data,
  title={Data poisoning attack against recommender system using incomplete and perturbed data},
  author={Zhang, Hengtong and Tian, Changxin and Li, Yaliang and Su, Lu and Yang, Nan and Zhao, Wayne Xin and Gao, Jing},
  booktitle={Proceedings of the 27th ACM SIGKDD Conference on Knowledge Discovery \& Data Mining},
  pages={2154--2164},
  year={2021}
}

@inproceedings{rastegarpanah2019fighting,
  title={Fighting fire with fire: Using antidote data to improve polarization and fairness of recommender systems},
  author={Rastegarpanah, Bashir and Gummadi, Krishna P and Crovella, Mark},
  booktitle={Proceedings of the twelfth ACM international conference on web search and data mining},
  pages={231--239},
  year={2019}
}

@inproceedings{li2021user,
  title={User-oriented fairness in recommendation},
  author={Li, Yunqi and Chen, Hanxiong and Fu, Zuohui and Ge, Yingqiang and Zhang, Yongfeng},
  booktitle={Proceedings of the Web Conference 2021},
  pages={624--632},
  year={2021}
}

@article{li2023fairness,
  title={Fairness in Recommendation: Foundations, Methods and Applications},
  author={Li, Yunqi and Chen, Hanxiong and Xu, Shuyuan and Ge, Yingqiang and Tan, Juntao and Liu, Shuchang and Zhang, Yongfeng},
  journal={ACM Transactions on Intelligent Systems and Technology},
  year={2023},
  publisher={ACM New York, NY}
}

@inproceedings{zhu2018fairness,
  title={Fairness-aware tensor-based recommendation},
  author={Zhu, Ziwei and Hu, Xia and Caverlee, James},
  booktitle={Proceedings of the 27th ACM international conference on information and knowledge management},
  pages={1153--1162},
  year={2018}
}

@inproceedings{yanan2024attack,
  title={Security of Recommender System: Adversarial Attack, Vulnerability Estimation and Mitigation Practice},
  author={Wang, Yanan and Ge, Yong},
  booktitle={INFORMS Workshop on Data Science},
  year={2024}
}

@inproceedings{patro2020fairrec,
  title={Fairrec: Two-sided fairness for personalized recommendations in two-sided platforms},
  author={Patro, Gourab K and Biswas, Arpita and Ganguly, Niloy and Gummadi, Krishna P and Chakraborty, Abhijnan},
  booktitle={Proceedings of the web conference 2020},
  pages={1194--1204},
  year={2020}
}

@article{khaleel2024network,
  title={Network and cybersecurity applications of defense in adversarial attacks: A state-of-the-art using machine learning and deep learning methods},
  author={Khaleel, Yahya Layth and Habeeb, Mustafa Abdulfattah and Albahri, AS and Al-Quraishi, Tahsien and Albahri, OS and Alamoodi, AH},
  journal={Journal of Intelligent Systems},
  volume={33},
  number={1},
  pages={20240153},
  year={2024},
  publisher={De Gruyter}
}

@article{lee2012shilling,
  title={Shilling attack detection—a new approach for a trustworthy recommender system},
  author={Lee, Jong-Seok and Zhu, Dan},
  journal={INFORMS Journal on Computing},
  volume={24},
  number={1},
  pages={117--131},
  year={2012},
  publisher={INFORMS}
}

@article{huang2011does,
  title={Why does collaborative filtering work? transaction-based recommendation model validation and selection by analyzing bipartite random graphs},
  author={Huang, Zan and Zeng, Daniel Dajun},
  journal={INFORMS Journal on Computing},
  volume={23},
  number={1},
  pages={138--152},
  year={2011},
  publisher={INFORMS}
}

\begin{IEEEbiography}[
{\includegraphics[width=1in,height=1.25in,clip,keepaspectratio]{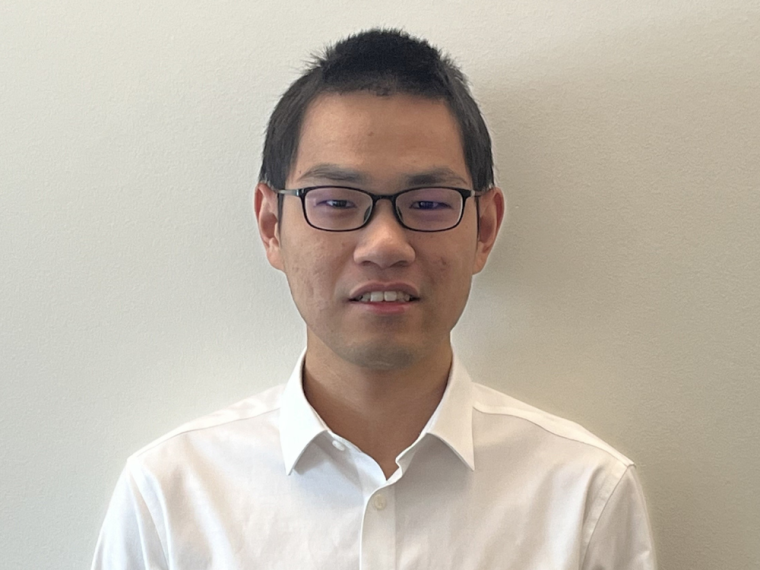}}]{\bf Yanan Wang} is currently an assistant professor in the Department of Information Systems and Operations Management, College of Business, The University of Texas at Arlington. He
received the PhD in Management Information Systems from University of Arizona in 2025. His main research interests include reinforcement learning, large language models, AI security and AI alignment.
\end{IEEEbiography}

\begin{IEEEbiography}[
{\includegraphics[width=1in,height=1.25in,clip,keepaspectratio]{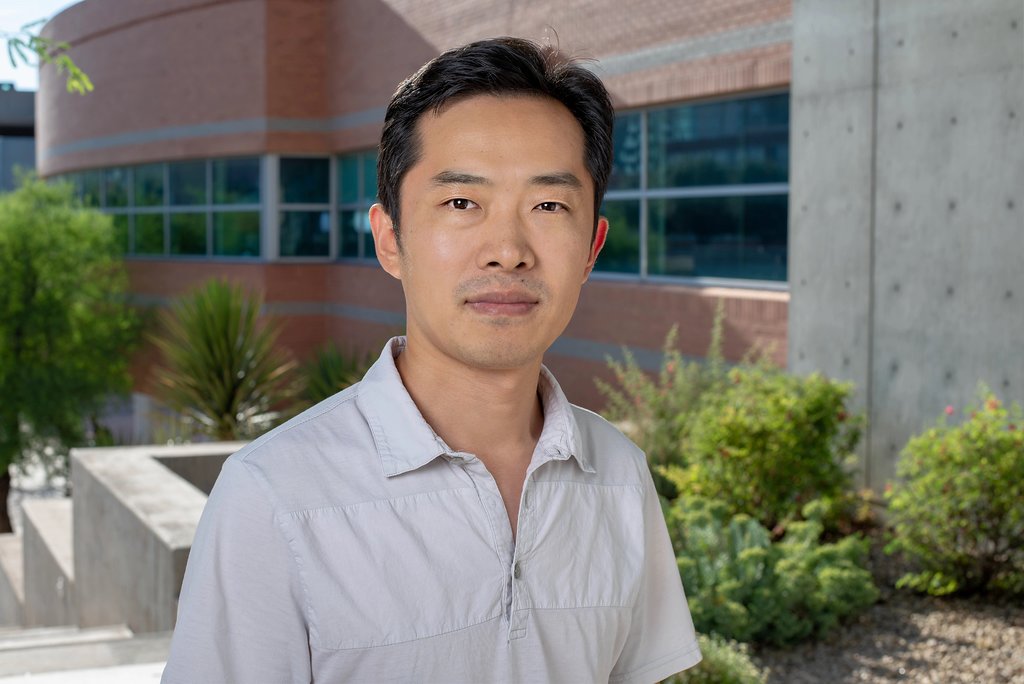}}] {\bf Yong Ge} is a professor of Management Information Systems at the Eller College of Management at the University of Arizona. He received his PhD in Information Technology from Rutgers Business School at Rutgers, The State University of New Jersey in 2013. His primary research interests include data mining, big data, machine/deep learning, recommender systems, personalization services, social networking, target marketing, talent analytics, and business analytics. 

\end{IEEEbiography}

\end{document}